\documentclass[pre,aps,twocolumn,superscriptaddress,showpacs]{revtex4}

\usepackage{amssymb,epsfig,amsmath,bm,subfigure,graphicx,textcomp,url,float,color,cases}

\begin{document}

\title{Dynamical phase transition in large deviation statistics of the Kardar-Parisi-Zhang equation}

\author{Michael Janas}
\email{jana0030@umn.edu}
\affiliation{Department of Physics, University of Minnesota, Minneapolis, MN 55455, USA}

\author{Alex Kamenev}
\email{kamenev@physics.umn.edu}
\affiliation{Department of Physics, University of Minnesota, Minneapolis, MN 55455, USA}
\affiliation{William I. Fine Theoretical Physics Institute, University of Minnesota,
Minneapolis, MN 55455, USA}

\author{Baruch Meerson}
\email{meerson@mail.huji.ac.il}
\affiliation{Racah Institute of Physics, Hebrew University of Jerusalem, Jerusalem 91904, Israel}


\begin{abstract}

We study the short-time behavior of the probability distribution $\mathcal{P}(H,t)$
of the surface height $h(x=0,t)=H$ in the
Kardar-Parisi-Zhang (KPZ) equation in $1+1$ dimension. The process starts from a stationary interface:
$h(x,t=0)$ is given by a realization of two-sided Brownian motion constrained by $h(0,0)=0$.
We find
a singularity  of the large deviation function of $H$
at a critical value $H=H_c$. The singularity has the character of a second-order phase transition. It reflects
spontaneous breaking of the reflection symmetry $x \leftrightarrow -x$ of  optimal  paths $h(x,t)$ predicted
by the weak-noise theory of the KPZ equation. At $|H|\gg |H_c|$ the corresponding tail of $\mathcal{P}(H)$ scales as $-\ln \mathcal{P} \sim |H|^{3/2}/t^{1/2}$ and
agrees, at any $t>0$, with the proper tail of the Baik-Rains distribution, previously observed only at long times.
The other tail
of $\mathcal{P}$ scales as $-\ln \mathcal{P} \sim |H|^{5/2}/t^{1/2}$ and coincides with the corresponding
tail for the sharp-wedge initial condition.

\end{abstract}

\maketitle

\section{Introduction}
\label{sec:Intro}

Large deviation functions of
nonequilibrium stochastic systems
can exhibit singularities, i.e. nonanalytic dependencies on the system parameters. In dynamical systems with a few degrees of freedom the singularities can be associated with the Lagrangian singularities of the underlying optimal fluctuational paths leading to a specified large deviation \cite{low1,low2,low3}. In extended macroscopic systems the nature of such singularities,  identified as nonequilibrium phase transitions \cite{Derrida,Touchette,MFTreview}, is not yet fully understood. So far several examples of such singularities  \cite{Bodineau,Gabrielli,Kafri} have been found in stochastic lattice gases: simple microscopic models of stochastic particle transport \cite{Spohn,Liggett,Kipnis}.

Here we uncover a nonanalytic behavior in a large-deviation function of the iconic Kardar-Parisi-Zhang (KPZ)  equation \cite{KPZ}.   This equation represents an important universality class of nonconserved surface growth  \cite{HHZ,Barabasi,Krug,Corwin,QS,HHT,S2016}, which is directly accessible in experiment  \cite{experiment1,experiment2}.
In $1+1$ dimension the KPZ equation,
\begin{equation}\label{KPZoriginal}
\partial_{t}h=\nu \partial^2_{x}h+(\lambda/2)\left(\partial_{x}h\right)^2+\sqrt{D}\,\xi(x,t),
\end{equation}
describes the evolution of the interface height  $h(x,t)$ driven by a Gaussian white noise  $\xi(x,t)$ with zero mean
and covariance
$\langle\xi(x_{1},t_{1})\xi(x_{2},t_{2})\rangle = \delta(x_{1}-x_{2})\delta(t_{1}-t_{2})$. Without loss of generality we will assume that $\lambda<0$ \cite{signlambda}.

An extensive body of work on the KPZ equation addressed the self-affine properties of the growing interface and the scaling  behavior of the
interface height at long times \cite{HHZ,Barabasi,Krug}. In $1+1$ dimension, the height fluctuations grow  as $t^{1/3}$, whereas the correlation length scales as $t^{2/3}$. These exponents are hallmarks of the KPZ universality class.

Recently the focus of interest in the KPZ equation in $1+1$ dimension shifted toward the complete
probability distribution ${\mathcal P}(H,T)$   of the interface height $h(0,T)-h(0,0)=H$ (in a proper moving frame \cite{displacement}) at a specified point $x=0$ and at \emph{any} specified time $t=T>0$. This distribution depends on the initial condition \cite{Corwin,QS,HHT,S2016}. One natural choice of the initial condition is a \emph{stationary} interface: an interface that has evolved for a long time prior to $t=0$.  Mathematically, it is described by a two-sided Brownian
interface pinned at $x=0$. In this case, in addition to averaging over realizations of the dynamic stochastic process, one has to average over all possible  initial pinned Brownian interfaces with diffusivity $\nu$. Imamura and Sasamoto \cite{IS} and Borodin \textit{et al.} \cite{Borodinetal} derived exact explicit
representations for  ${\mathcal P}(H,T)$ in terms of the Fredholm determinants.
They also showed that, in the long-time limit and for typical fluctuations, ${\mathcal P}$ converges to the Baik-Rains distribution \cite{BR} that is also encountered in the studies of the stationary totally
asymmetric simple exclusion process, polynuclear growth and last passage percolation \cite{Corwin}.

Here we will be mostly interested in {\emph short} times. As we show, at short times the interface height exhibits very interesting large-deviation properties. Instead of extracting the short-time asymptotics from the (quite complicated) exact representations \cite{IS,Borodinetal}, we will employ the weak noise theory (WNT) of the KPZ equation \cite{Fogedby1998,Fogedby1999,Fogedby2009,KK2007,KK2008,KK2009} which directly probes the early-time regime \cite{MKV,KMS}. In the framework of the WNT, $-\ln {\mathcal P}$ is proportional to the ``classical" action over the \emph{optimal path}: the most probable history $h(x,t)$ (a nonrandom function of $x$ and $t$) conditioned on the specified large deviation. A crucial signature of the stationary interface is the \textit{a priori} unknown optimal \emph{initial} height profile which is selected by the system out of a class of functions $h(x,0)$ carrying certain probabilistic weights and constrained by $h(0,0)=0$.

The central result of this paper is that at short times the optimal path and the optimal initial profile exhibit breaking of a reflection symmetry $x\leftrightarrow -x$ at a certain critical value $H=H_c$. This leads to a nonanalytic  behavior of the large deviation function of $H$ defined below.  This nonanalyticity exhibits all the characteristics of a mean-field-like second-order phase transition, where the role of the equilibrium free energy is played by the large deviation function of $H$. The nonanalyticity occurs in the negative (for our choice of $\lambda<0$) tail of ${\mathcal P}$. At $|H|\gg |H_c|$ this tail scales as $-\ln \mathcal{P} \sim |H|^{3/2}/T^{1/2}$   and agrees, at any $T>0$, with the corresponding tail of the Baik-Rains distribution \cite{BR}. The latter was  previously  derived \cite{IS,Borodinetal} only at long times. Here we show that it is applicable at
{\emph any} time $T>0$ for $H<0$ and $|H|\gg |H_c|$. We also find that the opposite, positive tail scales, at large $H$, as
$-\ln \mathcal{P} \sim H^{5/2}/T^{1/2}$. It coincides, in the leading order, with the corresponding tail for the \emph{sharp-wedge} initial condition \cite{DMRS,KMS}, and we provide the reason for this coincidence.

The rest of the paper is organized as follows. In Section \ref{sec:scaling} we present the WNT formulation of the problem. Section \ref{sec:small-H} deals with the limit of small $H$ which describes a Gaussian distribution of typical height fluctuations at short times.
Section \ref{sec:phase transition}  describes a numerical algorithm for solving the WNT equations  and presents numerical evidence for the symmetry-breaking transition. In Sections \ref{sec:negative H} and \ref{sec:positive H} we present analytical results for large negative and positive $H$, respectively. We summarize and discuss our results in Section \ref{sec:discussion}. Some of the technical details are relegated to three appendices.

\section{Weak noise theory}
\label{sec:scaling}

Let us rescale $t/T\to t$, $x/\sqrt{\nu T} \to x$, and $|\lambda| h/\nu\to h$. Equation~(\ref{KPZoriginal}) becomes
\begin{equation}\label{KPZrescaled}
\partial_{t}h=\partial^2_{x}h-(1/2) \left(\partial_{x}h\right)^2+\sqrt{\epsilon} \,\xi(x,t),
\end{equation}
where $\epsilon=D\lambda^2 \sqrt{T}/\nu^{5/2}$ is a dimensionless noise magnitude. We are interested in the probability density of observing $h(x=0,t=1)=H$, where $H$ is rescaled by $\nu/|\lambda|$,
under the condition that $h(x,0)$ is a two-sided Brownian interface with $\nu=1$ and $h(x=0,t=0)=0$. In the physical variables $\mathcal{P}(H,T)$ depends on two parameters $|\lambda| H /\nu$ and $\epsilon$.

The weak-noise theory  assumes that $\epsilon$ is a small parameter. The stochastic problem for Eq.~(\ref{KPZrescaled}) can be formulated as a functional integral which, in the limit of $\epsilon \ll 1$, admits a ``semi-classical'' saddle-point  evaluation. This  leads (see Appendix \ref{app:WNT})  to a minimization problem for the action functional $s=s_{\text{in}}+s_{\text{dyn}}$, where
\begin{eqnarray}\label{actn}
s_{\text{dyn}}&=& \frac{1}{2}\int_{0}^{1}dt\int_{-\infty}^{\infty} dx \left[\partial_{t} h-\partial_{x}^2 h+\frac{1}{2} \left(\partial_{x} h\right)^2\right]^2
\end{eqnarray}
is the dynamic contribution, and
\begin{equation}\label{s0}
s_{\text{in}} = \int_{-\infty}^{\infty}  dx \,(\partial_x h)^2|_{t=0}
\end{equation}
is the ``cost" of the (\textit{a priori} unknown) initial height profile \cite{only1d}. The ensuing Euler-Lagrange equation can be cast into two Hamilton equations
for the optimal path $h(x, t)$ and the canonically conjugate ``momentum" density $\rho(x,t)$:
\begin{eqnarray}
  \partial_{t} h &=& \delta \mathcal{H}/\delta \rho = \partial_{x}^2 h -(1/2) \left(\partial_x h\right)^2+\rho ,  \label{eqh}\\
  \partial_{t}\rho &=& - \delta \mathcal{H}/\delta h = - \partial_{x}^2 \rho - \partial_x \left(\rho \partial_x h\right) ,\label{eqrho}
\end{eqnarray}
where
$$
\mathcal{H} = \int dx \rho\left[\partial_x^2 h-(1/2) \left(\partial_x h\right)^2+\rho/2\right]
$$
is the Hamiltonian. Equations (\ref{eqh}) and (\ref{eqrho}) were first obtained by Fogedby \cite{Fogedby1998}.

Specifics of the one-point height statistics are reflected in the boundary conditions. The condition  $h(x=0,t=1)=H$ leads to
\cite{KK2007,MKV}
\begin{equation}\label{pT}
    \rho(x,t=1)=\Lambda \,\delta(x),
\end{equation}
where $\Lambda$ should be ultimately expressed in terms of $H$. The initial condition for the stationary interface follows from the variation of the action functional $s$ over $h(x,t=0)$ \cite{DG}, see Appendix \ref{app:WNT}, and takes
the form \cite{foot}
\begin{equation}
 \rho(x,t=0)+2\partial_x^2 h(x,t=0) = \Lambda \delta(x). \label{BC0}
\end{equation}
To guarantee the boundedness of the action, $\rho(x,t)$ and $\partial_x h(x,0)$ must go to zero sufficiently rapidly at $|x|\to \infty$. Finally,
\begin{equation}\label{pinned}
h(x=0,t=0)=0.
\end{equation}
Once the optimal path is found, we can evaluate $s=s_{\text{in}}+s_{\text{dyn}}$, where $s_{\text{dyn}}$ can be recast as
\begin{equation}
s_{\text{dyn}} =\frac{1}{2}\int_0^1 dt \int_{-\infty}^{\infty}   dx\,\rho^2 (x,t). \label{action1}
\end{equation}
This yields ${\mathcal P}$ up to pre-exponential factors: $-\ln \mathcal{P}\simeq s/\epsilon$. In the physical variables
\begin{equation}\label{actiondgen}
-\ln \mathcal{P}(H,T)\simeq \frac{\nu^{5/2}}{D\lambda^2\sqrt{T}}\,\,
s\left(\frac{|\lambda| H}{\nu}\right).
\end{equation}
As one can see, the action $s$ plays the role of the large deviation function for the short-time one-point height distribution. Below we determine the optimal path and $s$ analytically in different limits, and also evaluate these quantities numerically.

\vspace{0.5 cm}

\section{Small-$H$ expansion}
\label{sec:small-H}

\begin{figure}
\includegraphics[width=0.43\textwidth,clip=]{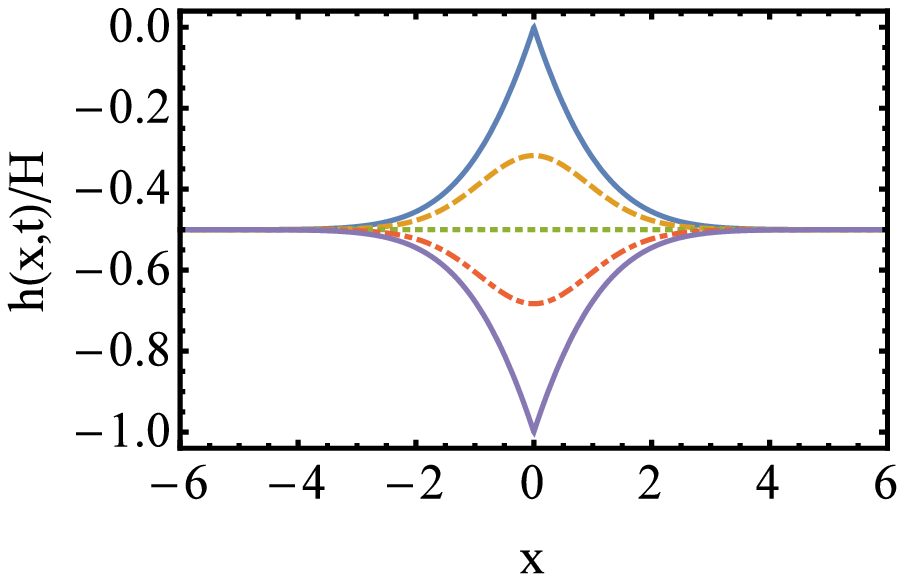}
\includegraphics[width=0.43\textwidth,clip=]{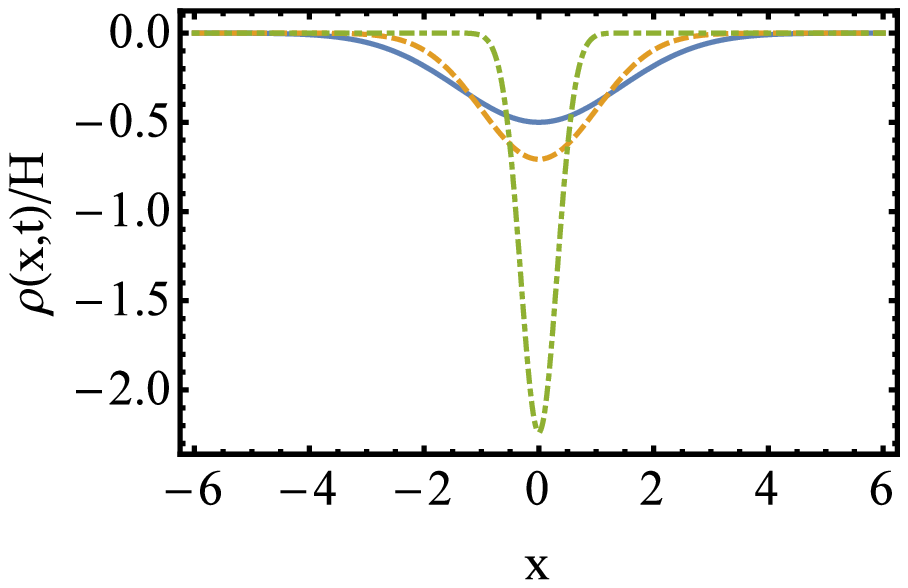}
\caption{The optimal path in the linear approximation, as described by Eqs.~(\ref{EWhistoryequation}) and (\ref{EWrho}) for $H<0$, at  $t=0$, $0.25$, $0.5$,
$0.75$ and $1$ (top panel, from top to bottom) and $t=0$, $0.5$ and $0.95$ (bottom panel).}
\label{EWhistory}
\end{figure}

For sufficiently small $H$ the WNT problem can be solved via a regular
perturbation expansion in the powers of $H$, or $\Lambda$ \cite{MKV,KMS,KrMe}. One writes
$h(x,t)= \Lambda h_1(x,t)+\Lambda^2 h_2(x,t) +\dots$  and similarly for $\rho(x,t)$, and obtains an iterative set of coupled {\emph linear} partial differential equations for  $h_i$ and $\rho_i$. These equations can be solved order by order with the standard Green function technique \cite{MKV}. The leading order corresponds to the WNT of the Edwards-Wilkinson equation \cite{EWpaper}:
\begin{eqnarray}
  \partial_t h_1 &=& \partial_x^2 h_1 +\rho_1,\label{heqEW0}\\
  \partial_t \rho_1 &=& -\partial_x^2 \rho_1, \label{peqEW0}
\end{eqnarray}
with the boundary conditions $\rho_1(x,0)+2\partial_x^2 h_1(x,0) = \rho(x,1) = \delta(x)$ and $h_1(0,0)=0$.
This is a simple problem, and one obtains in this order $\Lambda\simeq \sqrt{\pi}H$, and
\begin{eqnarray}
 \!\!h(x,t)\!&\simeq&\!\frac{H}{4}\left[2+xf\left(\frac{x}{2\sqrt{t}}\right)-xf\left(\frac{x}{2\sqrt{1-t}}\right)\right], \label{EWhistoryequation}\\
  \!\!\rho(x,t)\!&\simeq\!&\frac{H}{2\sqrt{1-t}}\,e^{-\frac{x^2}{4(1-t)}},  \label{EWrho}
\end{eqnarray}
where $f(z) = \sqrt{\pi}\,\text{erf}(z)+z^{-1}e^{-z^2}$, see Fig. \ref{EWhistory}. Noticeable in Eq.~(\ref{EWhistoryequation}) is a time-independent plateau  $h(\pm\infty,t)=H/2$.
Importantly for the following, $h(x,t)$ and $\rho(x,t)$ are, at all times,  symmetric functions of $x$. Although the KPZ nonlinearity appears already in the second order of the perturbation theory, the reflection symmetry
$x\leftrightarrow -x$  of the optimal path
persists in \emph{all} orders. Therefore,  within its (\textit{a priori} unknown) convergence radius, the perturbation series for $s(H)$ comes from  a unique optimal path which respects the reflection symmetry. Note for comparison that the \emph{time-reversal} symmetry $t\leftrightarrow 1-t$ of $h(x,t)$, present in the first order in $H$, is violated already in the second order, reflecting the lack of detailed balance in the KPZ equation.

Using Eqs.~(\ref{actn}) and~(\ref{s0}), one obtains, in the first order, $s(H)\simeq \sqrt{\pi} H^2/2$.  Therefore, as is well known, the body of the short-time distribution  $\mathcal{P}(H,T)$ is a Gaussian with the variance
$(D^2 T/\pi \nu)^{1/4}$ that obeys the Edwards-Wilkinson scaling \cite{EWpaper}. This variance is larger by a factor $\sqrt{2}$ than the variance for a flat initial interface, as observed long ago  \cite{Krugetal}. Indeed, a flat interface is  \emph{not} the optimal initial configuration for the stationary process, see
Fig.~\ref{EWhistory}.

\section{Phase transition at $H<0$: numerical evidence}
\label{sec:phase transition}

To deal with finite $H$ we used a numerical iteration algorithm \cite{Chernykh,EK} which cyclically solves Eq.~(\ref{eqrho}) backward in time, and Eq.~(\ref{eqh}) forward in time,  with the initial conditions (\ref{pT}) and (\ref{BC0}), respectively. At the very first iteration of Eq.~(\ref{eqrho}) one chooses a reasonable ``seed" function for $h(x,t)$ and keeps
iterating until the algorithm converges.
For small $|H|$ we used the linear theory, described above, to choose such a seed. We then used  $h(x,t)$, obtained upon convergence of the algorithm for a given $H$, as a seed for a slightly larger value $H$, {\it etc}.

For sufficiently small $|H|$ the algorithm converges to a reflection-symmetric optimal path resembling (or, for still smaller $|H|$, almost coinciding with) the one shown in Fig.~\ref{EWhistory}. The reflection symmetry is also intact for {\emph any} positive $H$, although the optimal solution strongly deviates from the small-$H$ solution of Sec.~\ref{sec:small-H} once $H>1$.

\begin{figure}
\includegraphics[width=0.40\textwidth,clip=]{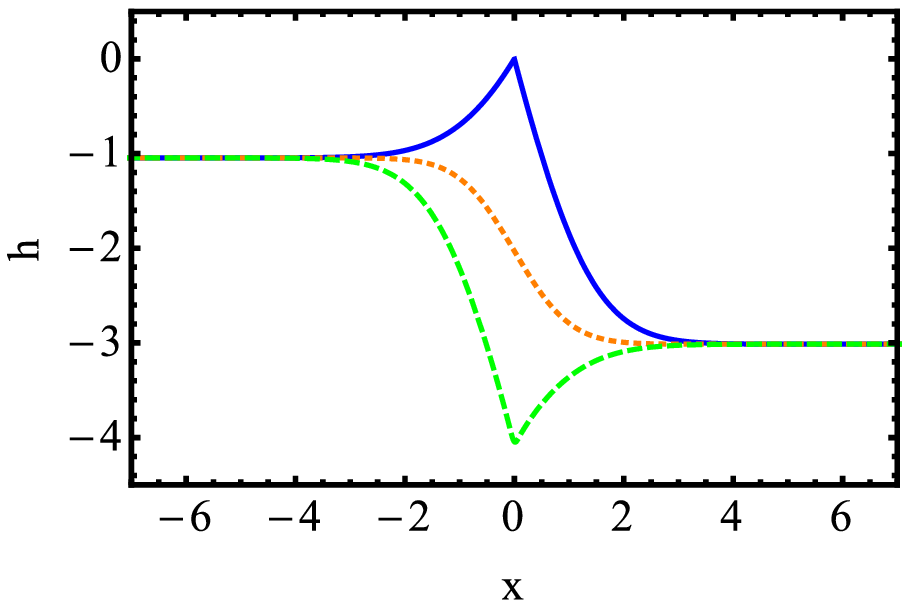}
\includegraphics[width=0.405\textwidth,clip=]{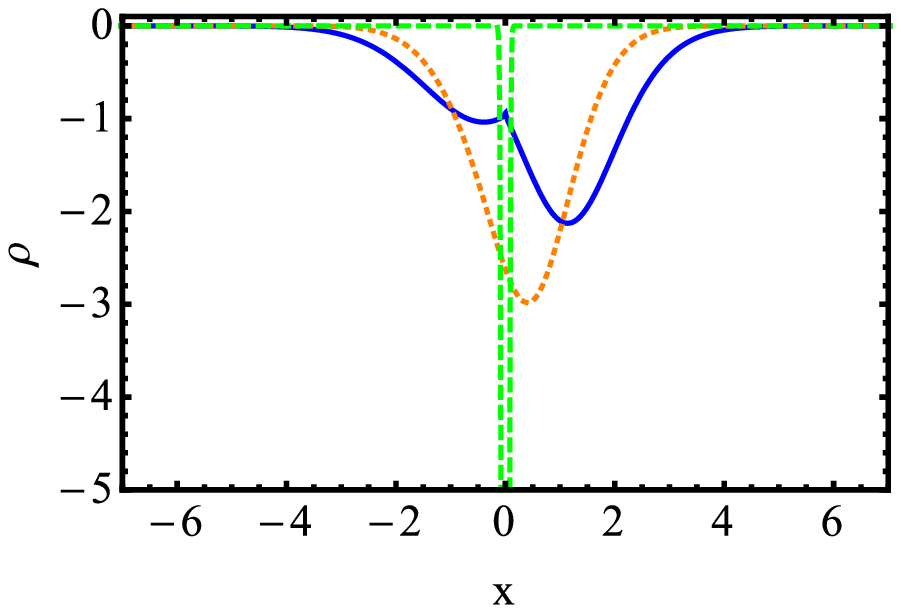}
\caption{The optimal path for $\Lambda=-6.3$ computed numerically. Shown are $h$ (top) and $\rho$ (bottom) vs. $x$ at $t=0$ (solid line), $0.5$ (short dash) and $1$ (long dash).}
\label{Lm6.3}
\end{figure}

\begin{figure}
\includegraphics[width=0.40\textwidth,clip=]{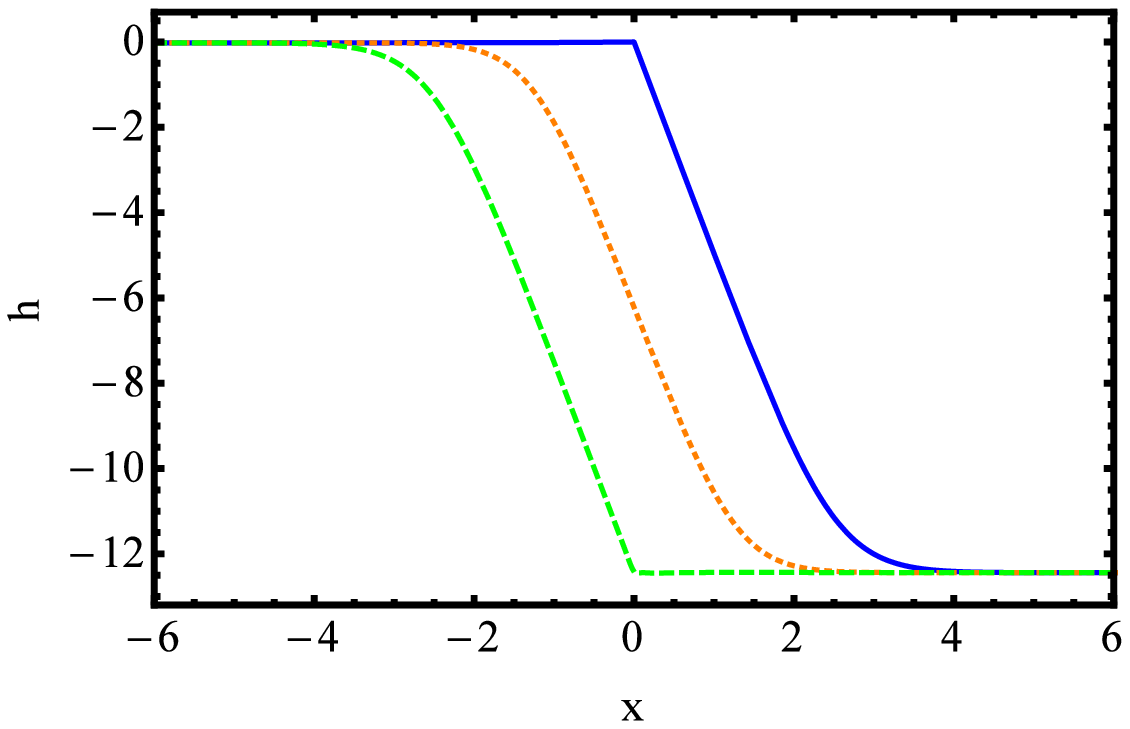}
\includegraphics[width=0.405\textwidth,clip=]{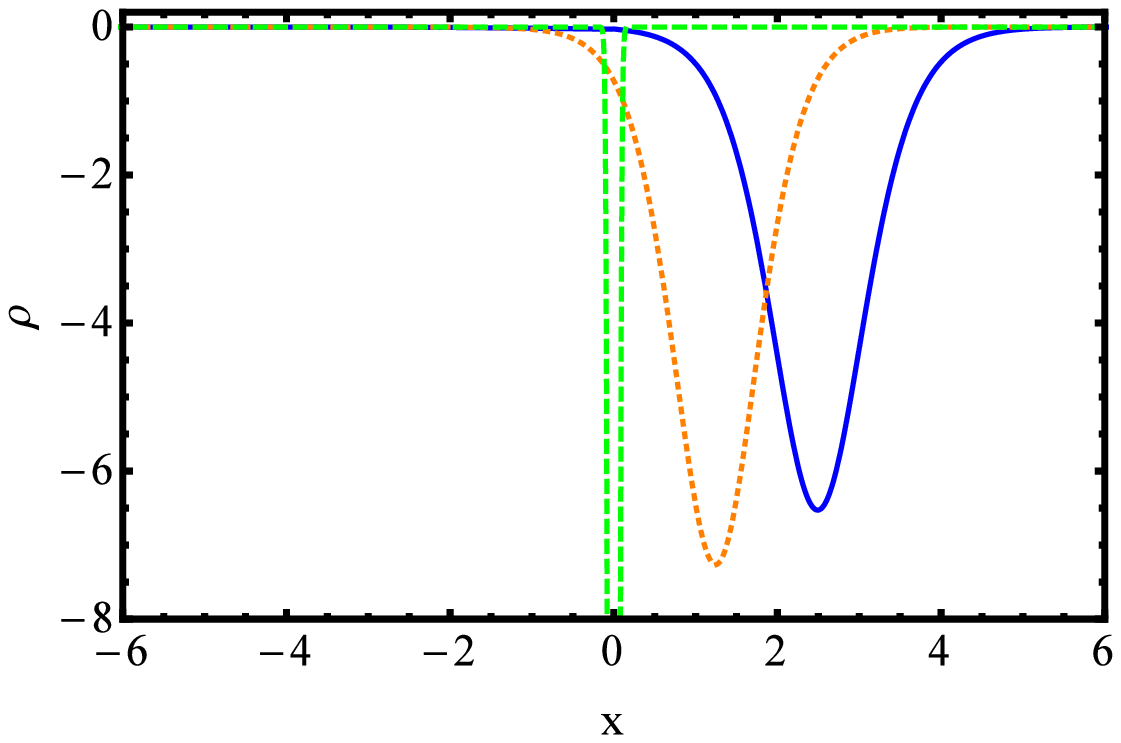}
\caption{Same as in Fig.~\ref{Lm6.3} but for $\Lambda=-10$.}
\label{Lm10}
\end{figure}

At sufficiently large {\emph negative} $H$ the symmetric solution loses
stability, and the algorithm converges  to one of two solutions with a {\emph broken} reflection symmetry. Each of these two solutions has unequal plateaus at $|x|\to \pm \infty$,  see Figs.~\ref{Lm6.3} and \ref{Lm10}, and is a  mirror reflection of the other around $x=0$.

\begin{figure}
\includegraphics[width=0.4\textwidth,clip=]{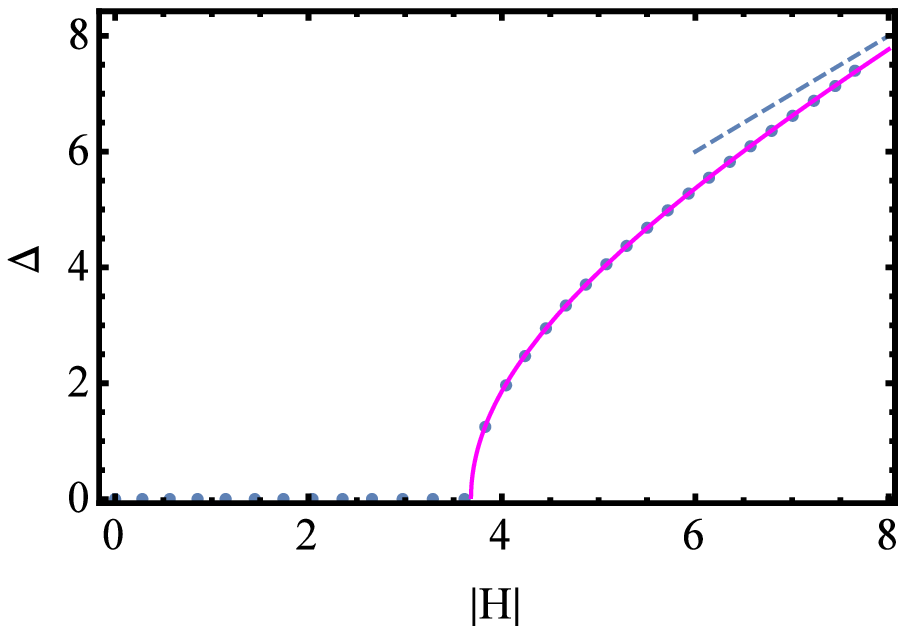}
\includegraphics[width=0.405\textwidth,clip=]{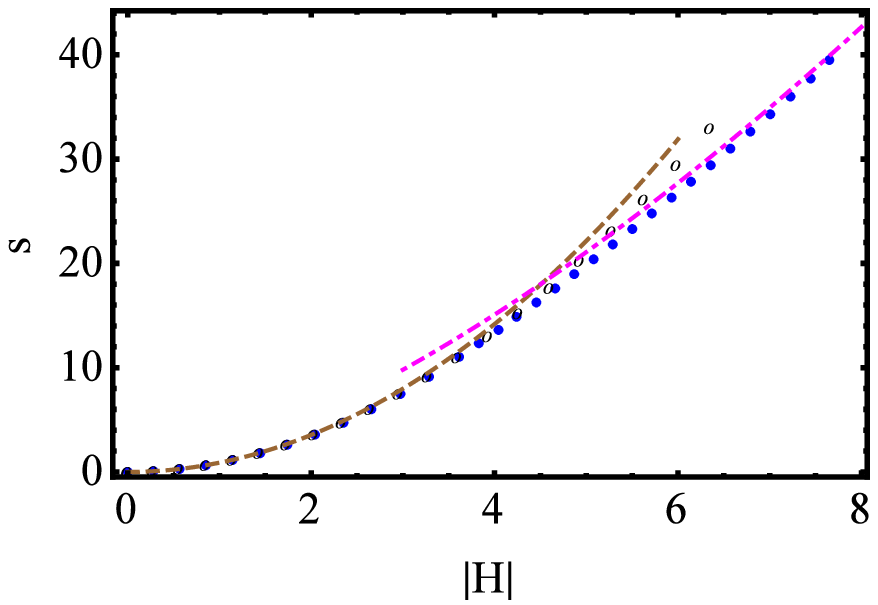}
\caption{Top:  $\Delta$ vs. $|H|$ at $H<0$. Symbols: numerical results, solid line: Eq.~(\ref{critical}), dashed line: the $|H|\gg 1$ asymptotic  $|\Delta| = |H|$.
Bottom:  $s$ vs. $|H|$ at $H<0$: The asymmetric and symmetric branches are shown by the solid and empty symbols, respectively. Also shown are the small- and large-$|H|$ asymptotics of $s$.}
\label{fig:Delta_and_s}
\end{figure}

To characterize the symmetry breaking
we introduced an order parameter
\begin{equation}
\label{delta}
\Delta=h(\infty,t)-h(-\infty,t)=\int_{-\infty}^{\infty} dx\, \partial_x h(x,t),
\end{equation}
which is a conserved quantity, as one can check from Eq.~(\ref{eqh}).
Our numerical results for $|\Delta|$ vs. $|H|$ at $H<0$ are shown in the top panel of Fig. \ref{fig:Delta_and_s}. They
indicate a phase transition at a critical value $H=H_c$. At  $|H|\leq|H_c|$ $\Delta=0$, in agreement with the results of the previous Section. For $|H|\geq|H_c|$ a good fit to the data is provided by
\begin{equation}\label{critical}
\Delta^2(H)=a(|H|-|H_c|)+b(|H|-|H_c|)^2,
\end{equation}
with $H_c\simeq -3.7$, $a\simeq 10.6$ and $b\simeq 0.8$. This suggests a mean-field-like second-order transition,
where the large deviation function  $s$ exhibits a discontinuity in its second derivative $\partial_H^2 s$ at $H=H_c$. One can recognize this discontinuity in the bottom panel of Fig. \ref{fig:Delta_and_s} which
shows $s$ vs. $H$ for the asymmetric (solid symbols) and symmetric (empty symbols) solutions \cite{artificial}.  The corresponding values of $s$ coincide
at $|H|<|H_c|$ but start deviating from each other at $|H|>|H_c|$, the symmetric solution becoming nonoptimal. The bottom panel also shows the small-$H$ analytic result $s(H)=\sqrt{\pi}H^2/2$, and the large-$|H|$ analytic result (\ref{sasym}) obtained below.

\section{Negative-$H$ tail}
\label{sec:negative H}

\begin{figure}
\includegraphics[width=0.40\textwidth,clip=]{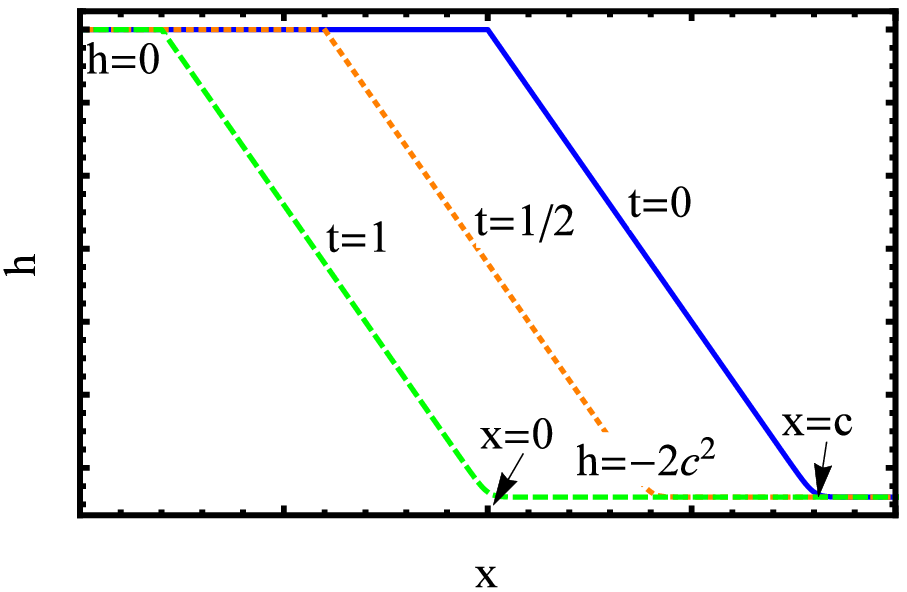}
\includegraphics[width=0.40\textwidth,clip=]{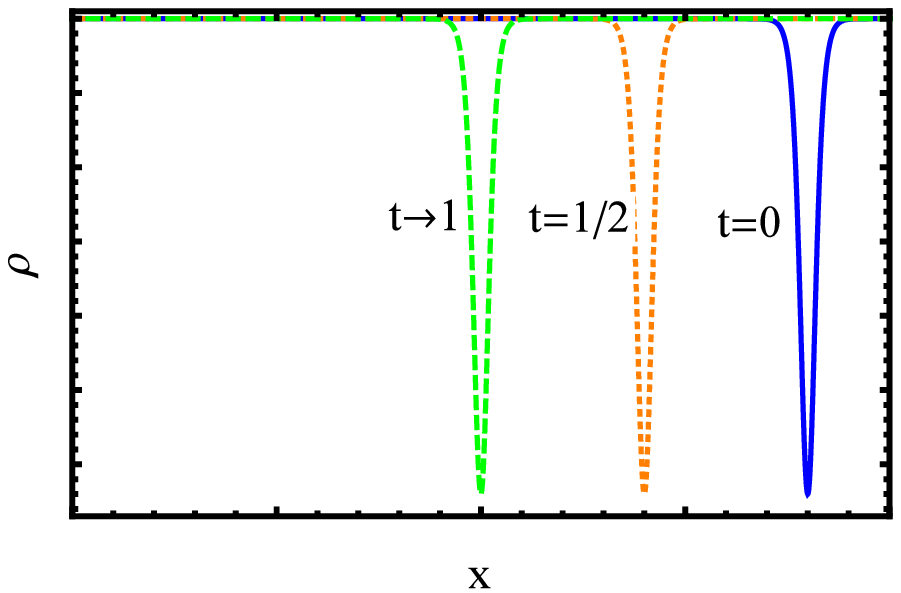}
\caption{The optimal path, $h(x,t)$ and $\rho(x,t)$, for $H<0$ and $|H|\gg 1$, see
Eqs.~(\ref{rhosoliton10})-(\ref{out}), for $t=0$, $1/2$ and $1$.}
\label{asymh}
\end{figure}

At very large negative $H$, or $\Lambda$, the asymmetric and symmetric solutions can be approximately found analytically. They involve narrow pulses of $\rho$, which we will call solitons, and ``ramps" of $h$. The asymmetric solutions can be parameterized by the soliton and ramp speed $c\gg 1$. The left-moving solution can be written as
\begin{eqnarray}
    \rho(x,t) &=&  -c^2 \,\text{sech}^2\left[\frac{c}{2}
  \, (c t+x-c)\right], \label{rhosoliton10}\\
  h(x,t)&\simeq& 2 \ln  \left[1+e^{c(ct+x-c)}\right] -2 c (c t+x) \label{hin10}
\end{eqnarray}
for $x>-ct$, and
\begin{equation}\label{out}
\rho(x,t)\simeq h(x,t)\simeq 0
\end{equation}
at $x<-ct$, see Fig.~\ref{asymh}. The expressions for each of the two regions are exact solutions of Eqs.~\ref{eqh}
and (\ref{eqrho}). The approximate combined solution obeys, up to exponentially small corrections, the boundary conditions~(\ref{BC0}) and  (\ref{pinned}).  It is continuous (again, up to an exponentially small correction), but  includes a shock in the interface slope $V(x,t)=\partial_x h(x,t)$ at $x=-ct$ \cite{matching}.   In our numerical solutions for large negative $\Lambda$, the $\rho$-soliton rapidly changes into the delta-function~(\ref{pT}) at $t\to 1$ (as Fig.~\ref{Lm10} indicates already for moderate negative $\Lambda$).  This transient does not contribute to the action in the leading order in $|H|\gg 1$.

The conservation law $\int dx\,\rho(x,t)=\Lambda$  yields $c=-\Lambda/4$, and we obtain $s=s_{\text{dyn}}+s_{\text{in}}= 4c^3/3+4 c^3=(16/3)c^3$.
Expressing $c$ via $H$ from the relation $|H|=2 c^2$ (see Fig. \ref{asymh}), we arrive at
\begin{equation}\label{sasym}
s=\frac{4\sqrt{2} \,|H|^{3/2}}{3}.
\end{equation}
In the physical units
\begin{equation}\label{newtail}
-\ln \mathcal{P} (H,T)\simeq \frac{4\sqrt{2}\,\nu |H|^{3/2}}{3 D |\lambda|^{1/2} T^{1/2}},
\end{equation}
in perfect agreement with the proper tail
of the Baik-Rains distribution \cite{BR,units}. The latter has been known to describe the \emph{late-time} one-point statistics of the KPZ interface for the stationary initial condition \cite{IS,Borodinetal}. As we see now, this tail holds at any $T>0$.

The simplest among the \emph{symmetric} solutions is a single stationary $\rho$-soliton and two outgoing $h$-ramps. These exact solutions were found earlier \cite{KK2007,MKV,KMS}. A family of more complicated exact two-soliton solutions involves two counter-propagating $\rho$-solitons that collide and merge into a single stationary soliton. Correspondingly, two counter-propagating $h$-ramps disappear  upon collision and reemerge with the opposite  signs, see Appendix  \ref{app:Cole-Hopf}. Remarkably, the singe-soliton and two-soliton solutions are particular members of a whole family of exact multi-soliton/multi-ramp solutions of Eqs.~(\ref{eqh}) and (\ref{eqrho}). We found them by performing the Cole-Hopf canonical transformation $Q=e^{-\frac{h}{2}},\quad P=-2\rho\,e^{\frac{h}{2}}$ and applying the Hirota method \cite{Hirota} to the transformed equations; see  Appendix  \ref{app:Cole-Hopf} for more details.

For all symmetric solutions with $c\gg 1$ the action $s$ is twice as large as what Eq.~(\ref{sasym}) predicts, so they are not optimal.  Notably, the corresponding nonoptimal action $s$ coincides with that describing the tail of the Tracy-Widom distribution \cite{TW}. This tail appears, at all times, for a class of \emph{deterministic} initial conditions \cite{MKV,DMRS,KMS}. Therefore, fluctuations in the initial condition, intrinsic for the stationary interface,  greatly enhance (by the factor of $2$ in a large exponent) the negative tail of ${\mathcal P}(H)$.

\section{Positive-$H$ tail}
\label{sec:positive H}

The opposite tail is of a very different nature. In particular, the optimal solution maintains reflection symmetry at any positive $H$. In the spirit of Refs. \cite{KK2009,MKV,KMS} the leading-order solution at $H\gg 1$ can be obtained  in terms of  ``inviscid hydrodynamics"  which  neglects the diffusion terms in Eqs.~(\ref{eqh}), (\ref{eqrho}) and (\ref{BC0}).
The resulting equations describe expansion of a ``gas cloud" of density $\rho(x,t)$ and mass $\Lambda$ from the origin, followed by collapse back to the origin at $t=1$. The same flow appears for the (deterministic) sharp-wedge initial condition \cite{KMS}. Its exact solution is given in terms of a uniform-strain flow with compact support $|x|\leq \ell(t)$, see Appendix \ref{app:shocks}. Both $h(x,t)$ and $\rho(x,t)$ are symmetric with respect to the origin. This leads to $s\simeq s_{\text{dyn}} = 4\sqrt{2} H^{5/2}/(15 \pi)$ \cite{KMS}, in agreement with Ref. \cite{DMRS}, where the same short-time asymptotic was derived from the exact representation for ${\mathcal P}(H,T)$ for the sharp wedge \cite{Corwin,SS,Calabreseetal,Dotsenko2010,Amir}. In the physical units
\begin{equation}\label{inviscid}
-\ln \mathcal{P}(H,T)\simeq \frac{4\sqrt{2 |\lambda|}}{15 \pi D}\, \frac{H^{5/2}}{T^{1/2}}.
\end{equation}
This tail is governed by the KPZ nonlinearity and does not depend on $\nu$.  At $|x|>\ell(t)$  $\rho=0$,
and $V(x,t)$ obeys the deterministic Hopf equation $\partial_t V + V \partial_x V=0$ and must be continuous at $|x|=\ell(t)$, as for the sharp wedge \cite{KMS}. Still, this Hopf flow
is different from its counterpart for the sharp wedge. Indeed, in the latter case
$V(|x|\to \infty,t)\simeq x/t$. For the stationary interface $V(|x|\to \infty,t)$ must vanish. This
condition
can be obeyed only if the Hopf flow involves two symmetric shocks where  $|V|$ drops from a finite value to zero: one shock at $x=x_s(t)>\ell(t)$, another at $x=-x_s(t)<-\ell(t)$. The shock dynamics are described in Appendix \ref{app:shocks}. A (symmetric) time-independent plateau,  $h(|x|\to \infty,t)\simeq H/2$, appears in this limit too.  The characteristic length scale of the solution is $\sim \Lambda^{1/3} \sim H^{1/2}$. As a result, $s_{\text{in}}$ from Eq.~(\ref{s0}) scales as  $H^{3/2}$. This is much less than $s_{\text{dyn}} \sim H^{5/2}$, justifying our neglect of the diffusion term in Eq.~(\ref{BC0}).

\section{Summary and Discussion}
\label{sec:discussion}

We have determined the tails of the short-time interface-height distribution in the KPZ equation when starting from a stationary interface. As we have shown, the $|H|^{3/2}/T^{1/2}$ tail of the Baik-Rains distribution, earlier predicted for long times,  holds at all times. We argue (see also Refs. \cite{MKV,KMS}) that the other tail, $|H|^{5/2}/T^{1/2}$, also holds at long times once the condition $|H|\gg T$ is met. It would be interesting to derive this tail from
the exact representation \cite{IS,Borodinetal}.

A central result of this paper is the discovery of a dynamical phase transition in  the large deviation function of $H$ at $T \to 0$. The transition  occurs at $H=H_c \simeq 3.7 \,\nu/\lambda$ and is caused by a spontaneous breaking of the reflection symmetry $x \leftrightarrow -x$ of the optimal path responsible for a given $H$.
We provided numerical evidence that the transition is of the second order. Strictly speaking, the WNT predicts a true phase transition only at a single point $(H_c, 0)$ of the phase diagram $(H,T)$. At finite but short times the transition is smooth but sharp around $H_c$, and this sharp feature should be observable in stochastic simulations of the KPZ equation.  One can characterize the transition by measuring the probability distribution of $\Delta$ (a random quantity) \cite{foot-Delta} at fixed $H$. This distribution is expected to change, in the vicinity of the critical value $H_c<0$, from unimodal, centered at zero, to bimodal. At very large $|H|$ the positions of bimodality peaks should approach $\Delta \simeq \pm H$.

Can symmetry breaking of this nature be observed  for discrete models which belong to the KPZ universality class (defined by \emph{typical} fluctuations at long times)? One  natural lattice-model candidate is the Weakly Asymmetric Exclusion Process (WASEP) with random initial conditions drawn from the stationary measure. Not only does the WASEP  belong to the KPZ universality class, but it also exhibits the Edwards-Wilkinson dynamics at intermediate times: when the microscopic details of the model are already forgotten but the process is still in the weak-coupling regime \cite{Corwin}. Although short-time large deviations of the WASEP can be different from those of the KPZ equation, one can expect the symmetry-breaking phenomenon to be robust.

Finally, the dynamical phase transition reported here is a direct consequence of fluctuations in the initial condition.  Similar transitions, at the level of large deviation functions, may exist in other nonequilibrium systems, both discrete and continuous, which involve averaging over random initial conditions.

\section{Acknowledgments}

We thank V.E. Adler, J. Baik, E. Bettelheim and J. Krug for useful discussions. A.K. was supported by NSF grants DMR-1306734 and DMR-1608238. B.M. acknowledges support from the Israel Science
Foundation (Grant No. 807/16), from the United States-Israel Binational Science Foundation (BSF) (Grant No. 2012145),  and from the William I. Fine Theoretical Physics Institute of the University of Minnesota. A.K. and B.M. acknowledge support from the University of Cologne through the Center of Excellence ``Quantum Matter and Materials".

\appendix

\section{Derivation of the weak-noise equations and boundary conditions}
\label{app:WNT}

Using Eq.~(1), one can express the  noise term as
\begin{equation}\label{actn0}
\sqrt{D} \xi(x,t)=\partial_{t} h-\nu \partial_{x}^2 h-\frac{\lambda}{2} \left(\partial_{x} h\right)^2.
\end{equation}
The probability to encounter such a realization of the Gaussian white noise is given by  $\propto e^{-S_{\text{dyn}}/D}$, where
\begin{eqnarray}\label{actn1}
S_{\text{dyn}}&=&{D/2} \int_{0}^{T}dt\int dx\,\xi^{2}(x,t)\\
&=&\frac{1}{2}\int_{0}^{T}dt\int dx \left[\partial_{t} h-\nu \partial_{x}^2 h-\frac{\lambda}{2} \left(\partial_{x} h\right)^2\right]^2.
\nonumber
\end{eqnarray}
The cost of creating an (\textit{a priori} unknown)  initial interface profile is determined by the stationary height distribution of the KPZ equation:
$$
S_{\text{in}} = \nu \int dx \,(\partial_x h)^2|_{t=0}.
$$
For a weak noise and large deviations, the dominant contribution to the total action $S=S_{\text{dyn}}+S_{\text{in}}$ comes from the optimal path $h(x,t)$ that is found by minimizing $S$ with respect to all possible paths $h(x,t)$ obeying the boundary conditions. The variation of the total action is
\begin{eqnarray}\label{variation}
\delta S=&&\!\!\!\!\! \int_{0}^{T}\!\!\! dt\!\int\!\! dx\left[\partial_{t} h-\nu\partial_{x}^2 h-\frac{\lambda}{2} \left(\partial_{x} h\right)^2\right]\Big(\partial_t \delta h -\nu \partial_x^2 \delta h  \nonumber   \\
 &&-\lambda \partial_x h \,\partial_x \delta h\Big)
+2 \nu \int dx \,\partial_x h\,\partial_x \delta h|_{t=0}.
\end{eqnarray}
Let us introduce the momentum density field $\rho(x,t)=\delta L/\delta v$, where $v\equiv \partial_t h$, and
$$
L\{h\}=\frac{1}{2}\int dx \left[\partial_{t} h-\nu \partial_{x}^2 h-\frac{\lambda}{2} \left(\partial_{x} h\right)^2\right]^2
$$
is the Lagrangian.  We obtain
\begin{equation}\label{p}
\rho(x,t)=\partial_{t}h - \nu \partial_{x}^2 h -\frac{\lambda}{2} \left(\partial_x h\right)^2
\end{equation}
and arrive at
\begin{equation}\label{heq}
\partial_{t}h=\nu \partial_{x}^2 h +\frac{\lambda}{2} \left(\partial_x h\right)^2+\rho,
\end{equation}
the first of the two Hamilton equations of the weak-noise theory (WNT). Now we can rewrite the variation (\ref{variation}) as follows:
$$
\delta S=\int_{0}^{T}dt\int dx\,\rho \,(\partial_{t}\delta h-\nu\partial_{x}^2\delta h -\lambda \partial_x h \,\partial_x \delta h)
$$ $$ +2 \nu \int dx \,\partial_x h\,\partial_x \delta h|_{t=0}.
$$
Demanding $\delta S=0$ and performing  integrations by parts, one obtains the Euler-Lagrange equation,
which yields the second Hamilton equation of the WNT:
\begin{equation}\label{peq}
\partial_{t}\rho=-\nu \partial_{x}^2 \rho +\lambda \partial_x \left(\rho \partial_x h\right).
\end{equation}
The boundary terms in space, resulting from the integrations by parts, all vanish.
The boundary terms in time must vanish independently at $t=0$ and $t=T$.
Both $h(x,t=0)$, and $h(x,t=T)$  are arbitrary everywhere except at $x=0$ where they are fixed by the conditions
\begin{equation}\label{conditions}
h(x=0,t=0)=0 \quad \text{and} \quad h(x=0,t=T)=H.
\end{equation}
This leads to the following boundary conditions:
\begin{eqnarray}
 \rho(x,t=0)+2\nu \partial_x^2 h(x,t=0) &=& \Lambda \delta(x), \label{BC0phys}\\
 \rho(x,t=T) &=& \Lambda \delta(x), \label{BCTphys}
\end{eqnarray}
where $\Lambda$ is an auxiliary parameter that should be finally set by the second relation in Eq.~(\ref{BCTphys}).
An evident additional condition, $\partial_x h(|x|\to \infty,t)=0$, is necessary for the boundedness of $S_{\text{in}}$.
Once the WNT equations are solved, the desired probability density is given by
\begin{eqnarray}
 \label{action}
&&  -\ln \mathcal P(H,T) \simeq \frac{S}{D} \\
 && = \frac{1}{2D}\int_0^T dt \int dx\,\rho^2(x,t) + \frac{\nu}{D} \int dx \,(\partial_x h)^2|_{t=0}.
 \nonumber
\end{eqnarray}
The rescaling transformation
\begin{equation}\label{rescalingKPZ}
t/T\to t, \quad x/\sqrt{\nu T} \to x, \quad |\lambda| h/\nu\to h, \quad |\lambda| T p/\nu \to p
\end{equation}
brings Eqs.~(\ref{heq}) and (\ref{peq}) to the rescaled form (5) and (6) of the main text.
The boundary condition (\ref{BC0phys}) becomes Eq.~(8), with a rescaled $\Lambda$. The rest of boundary conditions
remain the same.

\renewcommand{\theequation}{B\arabic{equation}}
\setcounter{equation}{0}
\section{Cole-Hopf transformation, kinks, solitons and ramps}
\label{app:Cole-Hopf}

As explained in the main text, the optimal path at very large negative $H$ can be approximately described in terms
of a $\rho$-soliton and $h$-ramp. As we show here, this solution is a particular member
of a whole family of exact multi-soliton/multi-ramp solutions of the WNT equations.
Let us perform a canonical Cole-Hopf transformation from $h$ and $\rho$ to $Q$ and $P$ according to
\begin{equation}\label{Q}
Q=e^{-\frac{h}{2}},\quad P=-2\rho\,e^{\frac{h}{2}}.
\end{equation}
The inverse transformation is $h=-2 \ln Q$ and $\rho=-(1/2)\, QP$. 
In the new variables 
the Hamilton equations,
\begin{eqnarray}
  \partial_t Q &=& \partial_x^2 Q +\frac{1}{4} Q^2 P, \label{Qt}\\
  \partial_t P &=&  -\partial_x^2 P -\frac{1}{4} Q P^2, \label{Pt}
\end{eqnarray}
have a symmetric structure and appear 
in the \textit{Encyclopedia of Integrable Systems} \cite{encyclopedia}. In this work we do not pursue the complete
integrability aspects and limit ourselves to exact multi-kink solutions which we found using the Hirota method \cite{Hirota}.
The multi-kink solutions in terms of $Q$ and $P$ become multi-soliton and multi-ramp solutions in terms of $\rho$ and $h$, respectively. The Hirota ansatz
$$
Q=\frac{v}{u},\quad\quad P=\frac{w}{u},
$$
transforms Eqs.~(\ref{Qt}) and (\ref{Pt}) into  the following form:
\begin{eqnarray}
(D_t - D_x^2)(v\cdot u)&=&0 , \nonumber \\
(D_t + D_x^2)(w\cdot u)&=&0 , \label{eq:Hirota}\\
           D_x^2(u\cdot u)&=&\frac{1}{4}\,vw ,\nonumber
\end{eqnarray}
where $D_t(A\cdot B)=A_tB-AB_t$ and $D_x^2(A\cdot B)=A_{xx}B-2A_xB_x+AB_{xx}$ are the Hirota derivatives. Equations (\ref{eq:Hirota}) admit two families of $N$-kink solutions:
\begin{eqnarray}
u&=& \sum\limits_{i=1}^N \eta_i^{(+)}, \nonumber\\
v&=& \frac{1}{C}\,\sum\limits_{i,j=1}^N (c_i-c_j)^2 \eta_i^{(+)}  \eta_j^{(+)} ,\label{1}\\
w&=&4 C\nonumber,
\end{eqnarray}
and
\begin{eqnarray}
u&=&\sum\limits_{i=1}^N \eta_i^{(-)}, \nonumber\\
v&=&4 C, \label{2}\\
w&=& \frac{1}{C}\,\sum\limits_{i,j=1}^N (c_i-c_j)^2 \eta_i^{(-)}  \eta_j^{(-)}, \nonumber
\end{eqnarray}
where $\eta_i^{(\pm)}(x,t) = e^{\pm c_i^2t - c_i(x-X_i)}$, the kinks are parametrized by
$N$ velocities $c_i$ and $N$ initial coordinates $X_i$, $i=1,\ldots,N$, and $C$ is an arbitrary constant,
reflecting invariance of the original WNT equations (\ref{eqh}) and (\ref{eqrho}) with respect to an arbitrary shift of $h$.
For the family of solutions (\ref{1}) we obtain
\begin{eqnarray}
\!\!\!\!\!\!\!\!\!\!\!\!\!\!\!\!\!\!&&h(x,t) \nonumber \\
\!\!\!\!\!\!\!\!\!\!\!\!\!\!\!\!\!\!&&=2 \ln \left[
\frac{C \sum\limits_{i=1}^N e^{c_i  (c_it-x+X_i)}}{\sum\limits_{i,j=1}^N (c_i-c_j)^2 e^{c_i(c_i t-x+X_i)+c_j (c_j t  -x+ X_j)}}
\right], \label{h}\\
\!\!\!\!\!\!\!\!\!\!\!\!\!\!\!\!\!\!&&\rho(x,t) \nonumber \\
\!\!\!\!\!\!\!\!\!\!\!\!\!\!\!\!\!\!&&= -\frac{2 \sum\limits_{i,j=1}^N
   (c_i-c_j)^2 e^{c_i(c_i t-x+X_i)+c_j (c_j t
  -x+ X_j)}}{\left[\sum\limits_{i=1}^N e^{c_i
   (c_it-x+X_i)}\right]^2}.
   \label{rho}
\end{eqnarray}
\begin{figure}
\includegraphics[width=0.45\textwidth,clip=]{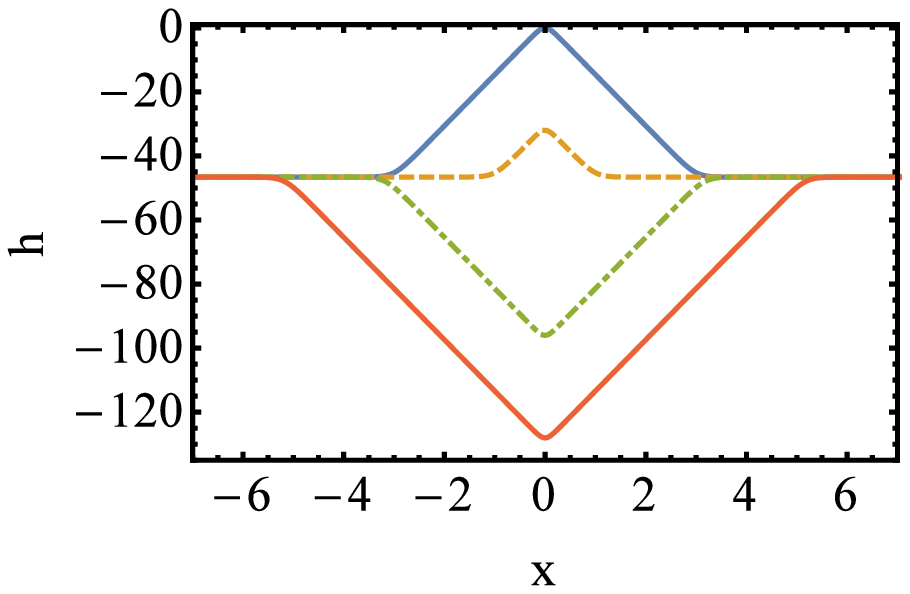}
\includegraphics[width=0.45\textwidth,clip=]{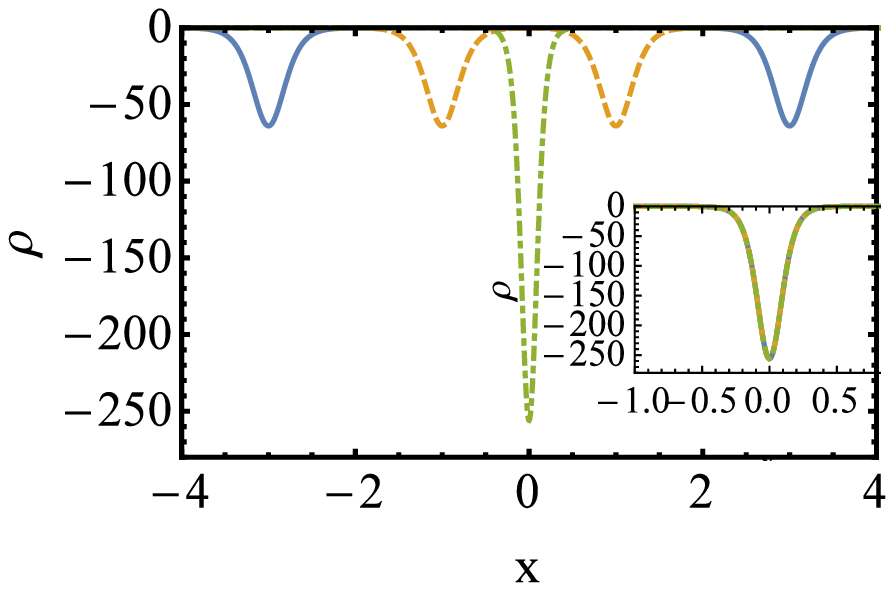}
\caption{Example of exact two-ramp/two-soliton solutions~(\ref{h}) and (\ref{rho}) for $h(x,t)$ and $\rho(x,t)$, respectively. Shown (for $c=8$) are $h$ and $\rho$ versus $x$ for $N=3, c_1=X_1=0, c_3=-c_2=c$ and $X_3=-X_2=-(3/8)c$. Top panel: $t=0$ (solid), $1/4$ (dashed), $3/4$ (dash-dotted) and $1$ (solid). Bottom panel:
 $t=0$ (solid), $1/4$ (dashed) and $1$ (dash-dotted). Inset: $\rho$ versus $x$ at  $t=1/2, 4/5$ and $1$. At $c\gg 1$ and $t>\tau$, $\rho$ approaches the exact stationary one-soliton solution \cite{Fogedby1998,KK2007,MKV}.}
\label{supplfig1}
\end{figure}
\begin{figure}
\includegraphics[width=0.45\textwidth,clip=]{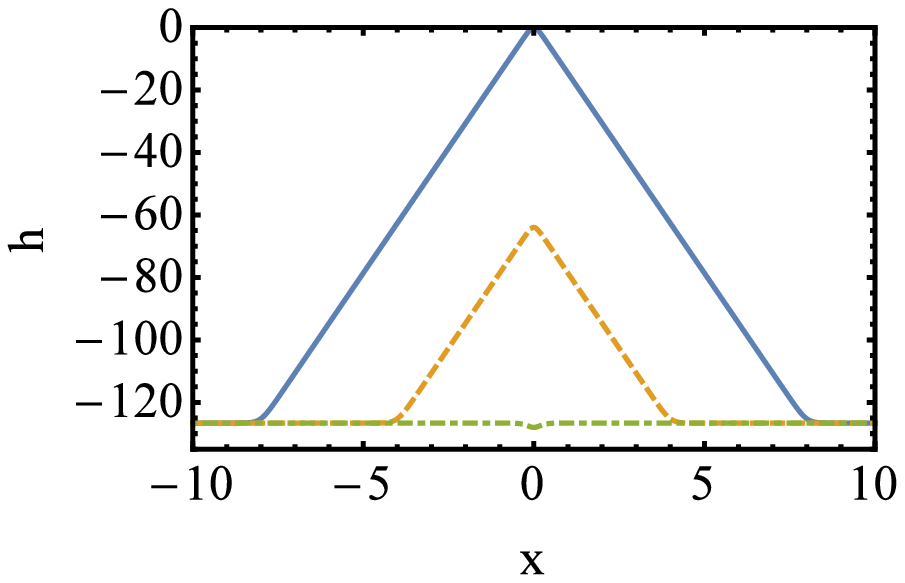}
\includegraphics[width=0.45\textwidth,clip=]{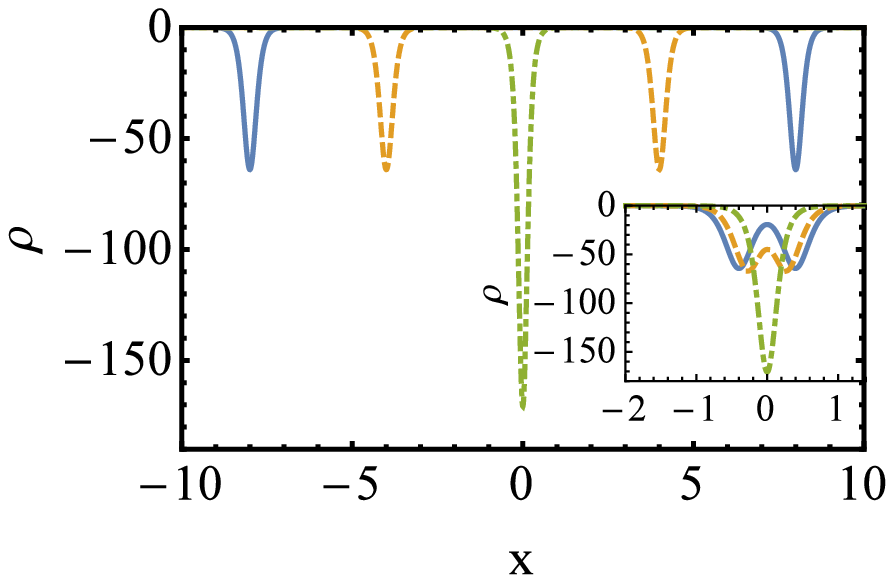}
\caption{Same as in Fig. \ref{supplfig1} but for $X_3=-X_2=-c$ (that is, $\tau=1$)
and $t=0$ (solid), $1/2$ (dashed) and $1$ (dash-dotted).  The inset shows a blowup of the collision and merger of the two  solitons at $t=0.95, 0.965$ and $1$.}
\label{supplfig2}
\end{figure}
The particular case of $N=3, c_1=X_1=0, c_3=-c_2=c$ and $X_3=-X_2=-c \tau$, where $0<\tau<1$, yields the family of symmetric solutions
described  in the context of large negative $H$ in Section \ref{sec:negative H}. Here two identical counter-propagating $\rho$-solitons collide and merge, at $x=0$, into a single soliton. The two ramps of $h$ also merge, but then change their signs and expand, see Figs. \ref{supplfig1} and \ref{supplfig2}.  At $c\gg 1$  these solutions approximately satisfy all the boundary conditions. The arbitrary constant $C$ can be chosen so as to impose the condition $h(x=0,t=0)=0$. However, for all these symmetric solutions (at fixed $c$ and different $\tau$) the total action $S$, in the leading order, is the same and \emph{twice as large} as $S$ for the asymmetric solution, described in the main text.  Therefore, neither of these solutions is optimal. Finally, the single stationary $\rho$-soliton, and the expanding ramps, observed at $t>\tau$  is by itself an exact solution of the WNT equations, as was previously known \cite{Fogedby1998,KK2007,MKV}. This solution corresponds to $\tau=0$ and represents the true optimal path for a whole class of \emph{deterministic} initial conditions \cite{KMS}.

\renewcommand{\theequation}{C\arabic{equation}}
\setcounter{equation}{0}
\section{Hydrodynamics and shocks for $H\gg 1$}
\label{app:shocks}

Here, in the spirit of Refs. \cite{KK2009,MKV}, the leading-order solution can be obtained  in terms of  ``inviscid hydrodynamics"  which  neglects the diffusion terms in Eqs.~(5), (6) and (8) of the main text.
The resulting equations for $\rho(x,t)$ and $V(x,t)=\partial_x h(x,t)$,
\begin{eqnarray}
 \partial_t \rho +\partial_x (\rho V)&=& 0, \label{rhoeq}\\
  \partial_t V +V \partial_x V &=&\partial_x \rho, \label{Veq}
\end{eqnarray}
describe expansion of a ``gas cloud" of density $\rho(x,t)$ and mass $\Lambda$ from the origin, followed by collapse back to the origin at $t=1$. The same flow appears for the (deterministic) sharp-wedge initial condition \cite{KMS}. Its exact solution is given in terms of a uniform-strain flow with compact support:
\begin{equation}
V(x,t)=-a(t)\,x, \quad |x|\leq \ell(t), \label{Vin}
\end{equation}
and
\begin{numcases}
{\!\!\rho(x,t) =}
r(t) \left[1-x^2/\ell^2(t)\right], & $|x|\leq \ell(t)$, \label{rhoin}\\
0, &$|x|> \ell(t)$. \label{rhoout}
\end{numcases}
As one can see, there is no symmetry breaking here. The functions $a(t)$, $\ell(t)$ and $r(t)$ were calculated in Ref. \cite{KMS}, leading to Eq.~(\ref{inviscid}).

\begin{figure}
\includegraphics[width=0.45\textwidth,clip=]{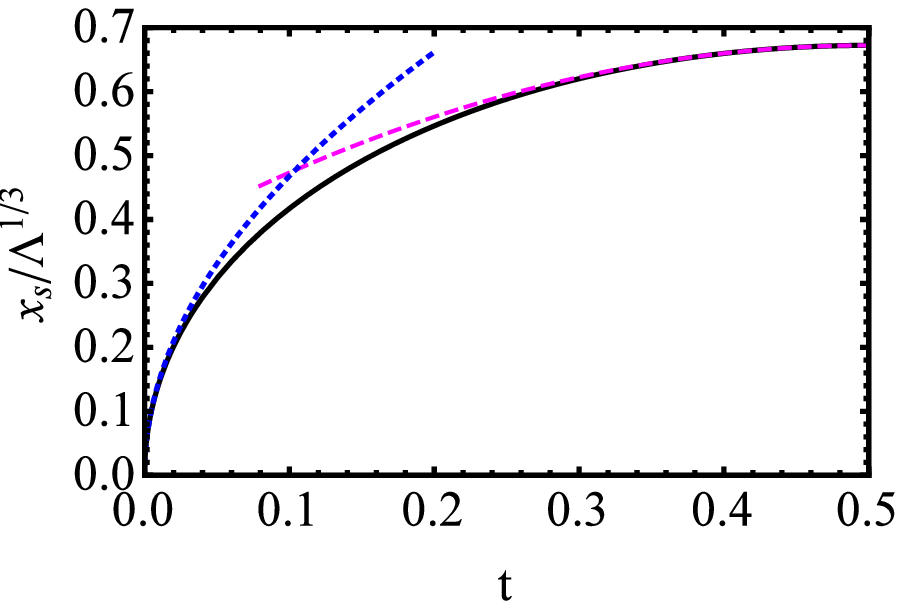}
\includegraphics[width=0.425\textwidth,clip=]{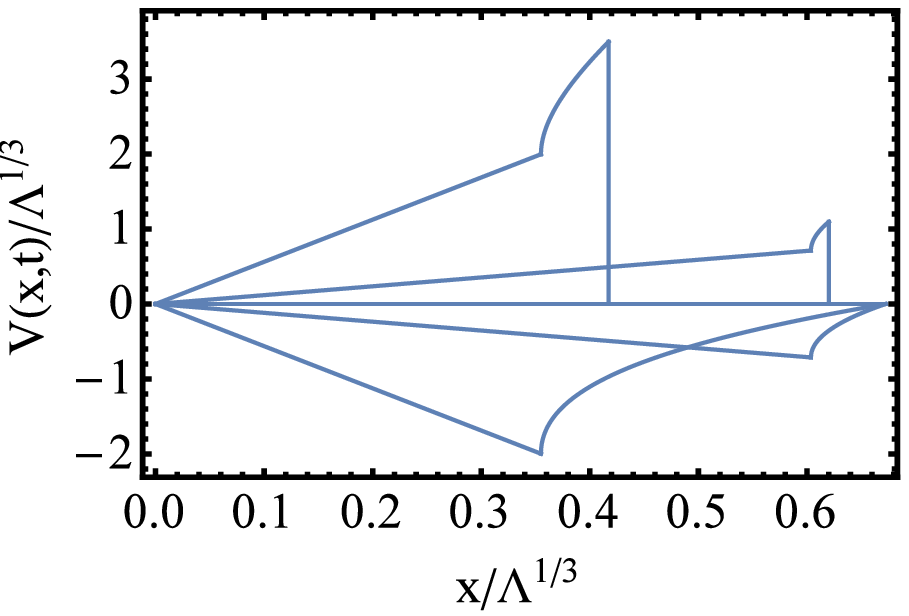}
\caption{Top: The shock position $x_s$ vs. time, alongside with its asymptotics (\ref{closeto05})  and~(\ref{closeto0}) for $x>0$. Bottom: $V=\partial_x h$ vs. $x$ at $H\gg 1$ at times (from top to bottom) $0.1$, $0.3$, $0.5$ (when $V=0$), $0.7$ and $0.9$. Both the uniform-strain solution~(\ref{Vin}),
and the Hopf solution with the shock are shown for $x>0$. In this limit $H=(3\pi \Lambda)^{2/3}/2$ \cite{KMS}.}
\label{fullflow}
\end{figure}

At $|x|>\ell(t)$ one has $\rho=0$. Here $V(x,t)$ obeys the deterministic Hopf equation $\partial_t V + V \partial_x V=0$ and must be continuous at $|x|=\ell(t)$ \cite{MKV,KMS}. In addition, we must demand $V(|x|\to \infty,t) = 0$. The latter
condition
can only be obeyed if the Hopf flow involves two symmetric shocks where  $|V|$ drops from a finite value to zero: one shock at $x=x_s(t)>\ell(t)$ (see the left panel of Fig.~\ref{fullflow}), another at $x=-x_s(t)<-\ell(t)$. The shocks are symmetric with respect to $x=0$, and their dynamics are quite interesting. Let us consider the $x>0$ shock. Its speed $\dot{x}_s$ must be equal to $(1/2) V[x_s(t)-0,t]$ \cite{Whitham}.  The expression for $V(x,t)$ can be found in Ref. \cite{KMS}. Upon rescaling $x$ and $V$ by $\Lambda^{1/3}$, one obtains the following differential equation for the shock position $x_s(t)$ at $x>0$:
\begin{equation}\label{ODEXs}
\dot{x}_s\left(1-2t -\frac{2}{\pi}\,\text{arctan}\,\frac{2\sqrt{\ell_0}\dot{x}_s}{\sqrt{3}}\right) = \ell_0-x_s,
\end{equation}
where $\ell_0=\ell(t=1/2)=3^{1/3}/\pi^{2/3}$ is the (rescaled) maximum size of the pressure-driven flow region. Equation~(\ref{ODEXs}) is of the first order but highly nonlinear. It should be solved on the time interval $0<t\leq 1/2$ with the initial condition $X_s(t=0)=0$. Close to $t=1/2$, when $x_s \to 0$ and $\dot{x}_s \to 0$,  we obtain a simple asymptotic:
\begin{equation}\label{closeto05}
\ell_0-x_s(t)\simeq \left(\frac{3\pi}{8}\right)^{5/3} \left(\frac{1}{2}-t\right)^2.
\end{equation}
At $t\to 0$  $x_s$ goes to zero and $\dot{X}_s$ goes to infinity. Expanding the arctangent at large argument up to and including the second term, we arrive at the linear equation
$2t \dot{x}_s = x_s$
and obtain the short-time asymptotic
\begin{equation}\label{closeto0}
x_s(t) \simeq K t^{1/2}
\end{equation}
with an unknown constant $K$ which can be found numerically. The shock magnitude (the jump of $V$) and speed decrease with time and vanish at $t=1/2$: the shocks disappear when they reach the stagnation points of the flow, $V=0$ which, according to Ref. \cite{KMS}, are located, at $t\geq 1/2$, at $x=\pm \ell(t=0)$. Notice that, at small $t$, $\ell (t)\sim t^{2/3}$, and the shock position is
indeed outside the pressure-driven region as we assumed. The left panel of Fig. \ref{fullflow} shows
the shock position $x_s(t)$ found by solving Eq.~(\ref{ODEXs}) numerically. Also shown are the asymptotic~(\ref{closeto05}), and
the asymptotic~(\ref{closeto0}) with $K=1.48$. The right panel of Fig. \ref{fullflow}  shows $V(x,t)$ vs. $x>0$ at different times.

Integrating $V(x,t)$ over $x$, one can obtain $h(x,t)$, but we do not show these
cumbersome formulas here.


\begin{thebibliography}{99}
\bibitem{signlambda}
Changing the sign of $\lambda$ is  equivalent to changing the sign of $h$.
\bibitem{displacement} The solution  of Eq.~(\ref{KPZoriginal}) includes a systematic interface displacement $h_s(x,t)$ that comes
from the noise rectification by
the nonlinearity \cite{Hairer,S2016}.
Our $H$ is defined as $H=h(0,t)-h_s(0,t)$.
\bibitem{only1d} The KPZ equation is one of a very few lucky nonequilibrium models
where the stationary height distribution, leading to Eq.~(\ref{s0}), is known (but only in one dimension).
\bibitem{foot} That $\Lambda$ in Eqs.~(\ref{pT}) and (\ref{BC0}) is the same follows from
the conservation law $\int_{-\infty}^{\infty} dx \,\rho(x,t)=\text{const}$ [cf. Eq.~(\ref{eqrho})]  and from the
boundary condition $\partial_x h(\pm \infty,0)=0$ that is necessary for $s_\text{in}$ to be bounded.
\bibitem{artificial} To obtain a converged symmetric solution, we artificially enforced symmetry at each iteration.
\bibitem{units}  To make the comparison, the
Baik-Rains distribution ${\mathcal P}_{\text{BR}}(H,T)$ should be represented as
$-\ln {\mathcal P}_{\text{BR}} (H,T) \simeq \Phi(H)/T^{1/2}$. The large deviation function
$\Phi(H)=s(-2H)/8$, where $s$ is our rescaled total action.
\bibitem{foot-Delta} In a stochastic system one can define $\Delta=h(X,0)-h(-X,0)$, where $X$ can be chosen as $a \ell$, where $\ell$ is the characteristic lateral size of the optimal initial height profile, predicted by the WNT at given $H$, and $a$ is a constant of order of unity. For large negative $H$,  $\ell \simeq (\lambda H T)^{1/2}$ is the lateral size of the ramp, see the left panels of Figs. \ref{Lm10} and \ref{asymh}.
\end{thebibliography}
\end{document}